\documentclass[%
reprint,
superscriptaddress,
showpacs,preprintnumbers,
amsmath,amssymb,
aps,
]{revtex4-1}

\usepackage{graphicx}
\usepackage{hyperref}% add hypertext capabilities

\begin{document}

%Title of paper
\title{Magnetic field direction dependent antiskyrmion motion with microwave electric fields}

\author{Chengkun Song}%
\affiliation{
	Key Laboratory for Magnetism and Magnetic Materials of the Ministry of Education, Lanzhou University, Lanzhou, 730000, People’s Republic of China
}%
%  \altaffiliation[Also at ]{Physics Department, XYZ University.}%Lines break automatically or can be forced with \\
\author{Chendong Jin}%
\affiliation{
	Key Laboratory for Magnetism and Magnetic Materials of the Ministry of Education, Lanzhou University, Lanzhou, 730000, People’s Republic of China
}%
%  \email{Second.Author@institution.edu}
\author{Haiyan Xia}%
\affiliation{
	Key Laboratory for Magnetism and Magnetic Materials of the Ministry of Education, Lanzhou University, Lanzhou, 730000, People’s Republic of China
}%
\author{Jinshuai Wang}%
\affiliation{
	Key Laboratory for Magnetism and Magnetic Materials of the Ministry of Education, Lanzhou University, Lanzhou, 730000, People’s Republic of China
}%
\author{Yunxu Ma}%
\affiliation{
	Key Laboratory for Magnetism and Magnetic Materials of the Ministry of Education, Lanzhou University, Lanzhou, 730000, People’s Republic of China
}%
\author{Jianbo Wang}%
\affiliation{
	Key Laboratory for Magnetism and Magnetic Materials of the Ministry of Education, Lanzhou University, Lanzhou, 730000, People’s Republic of China
}
\affiliation{
	Key Laboratory for Special Function Materials and Structural Design of the of Ministry of Education, Lanzhou University, Lanzhou, 730000, People's Republic of China
}%
\author{Qingfang Liu}
\email{liuqf@lzu.edu.edu}
\affiliation{
	Key Laboratory for Magnetism and Magnetic Materials of the Ministry of Education, Lanzhou University, Lanzhou, 730000, People’s Republic of China
}%

%Collaboration name if desired (requires use of superscriptaddress
%option in \documentclass). \noaffiliation is required (may also be
%used with the \author command).
%\collaboration can be followed by \email, \homepage, \thanks as well.
%\collaboration{}
%\noaffiliation

\date{\today}

\begin{abstract}
Magnetic skyrmions are regarded as promising information candidates in future spintronic devices, which have been investigated theoretically and experimentally in isotropic system. Recently, the stabilization of antiskyrmions in the presence of anisotropic Dzyaloshinskii-Moriya interaction and its dynamics driven by current have been investigated. Here, we report the antiskyrmion motion with the combined action of the in-plane magnetic field and microwave electric fields. The in-plane magnetic field breaks the rotation symmetry of the antiskyrmion, and perpendicular microwave electric field induces the pumping of magnetic anisotropy, leading to antiskyrmion breathing mode. With above two effects, the antiskyrmion propagates with a desired trajectory. Antiskyrmion propagation velocity depends on the frequency, amplitude of anisotropy pumping, and damping constant as well as strength of in-plane field, which reaches the maximum value when the frequency of microwave electric field is in consist with the resonance frequency of antiskyrmion. Moreover, we show that the antiskyrmion propagation depends on the direction of magnetic field, where the antiskyrmion Hall angle can be suppressed or enhanced. At a critical direction of magnetic field, the Hall angle is zero. Our results introduce a possible application of antiskyrmion in antiskyrmion-based spintronic devices with lower energy consumption.
\end{abstract}

% insert suggested PACS numbers in braces on next line
% insert suggested keywords - APS authors don't need to do this
%\keywords{}

%\maketitle must follow title, authors, abstract, \pacs, and \keywords
\maketitle

% body of paper here - Use proper section commands
% References should be done using the \cite, \ref, and \label commands
\section{Introduction}
Magnetic skyrmions are one of topological defects in low dimensional magnetic systems~\cite{nagaosa2013topological,romming2013writing,sampaio2013nucleation,fert2013skyrmions}. Compared to domain walls~\cite{parkin2008magnetic}, vortices~\cite{khvalkovskiy2009vortex} and magnetic bubbles~\cite{moutafis2009dynamics}, magnetic skyrmions exhibit topological protected stability~\cite{nagaosa2013topological,jiang2015blowing,woo2016observation}, and much attention has been focused on magnetic skyrmions. Depending on the Dzyaloshinskii-Moriya interaction (DMI) type, two types of skyrmions have been investigated, which are Bloch skyrmion in bulk DMI~\cite{jonietz2010spin, yu2011near,yu2012skyrmion,shibata2013towards,munzer2010skyrmion} and N$\acute{\mathrm{e}}$el skyrmions in the presence of interfacial DMI ~\cite{pollard2017observation,zhang2018direct,jiang2015blowing,jiang2017direct} . N$\acute{\mathrm{e}}$el skyrmions or Bloch skyrmions can be stabilized in materials belonging to crystallographic classes T (O) or $C_{nv}$~\cite{leonov2017skyrmion,bogdanov1989thermodynamically}, respectively. In the micromagnetic view, the interfacial and bulk DMI can be written as $D[\mathbf{m}(\nabla\cdot\mathbf{m})-(\mathbf{m}\cdot \nabla )\mathbf{m}]_z $ or $D\mathbf{m}\cdot(\nabla\times \mathbf{m})$, respectively~\cite{hoffmann2017antiskyrmions}, where skyrmions exhibit cylindrical symmetry in the isotropic environment. The magnetizations of domain wall (DW) in Bloch skyrmion is perpendicular to the radial direction, while the magnetization in N$\acute{\mathrm{e}}$el skyrmion is along the radial direction. 

Recently, magnetic antiskyrmions have been investigated that exist in anisotropic materials belonging to crystallographic classes $D_{2d}$ and $S_4$~\cite{hoffmann2017antiskyrmions,leonov2017skyrmion}, which break the cylindrical symmetry. Antiskyrmions have been theoretically predicted in bulk crystals~\cite{leonov2017skyrmion}, and recently they are also been demonstrated in an acentric tetragonal MnPtPdSn Heusler compound~\cite{nayak2017magnetic}. While antiskyrmions have not yet been discovered in thin film systems with interfacial DMI due to the reason that the thin film system leads to the same sign ($D_{x} = D_{y}$) and strength of DMI, they are unstable in 2D chiral magnets with scalar $D$. To stabilize antiskyrmions in thin film system with perpendicular anisotropy, the components of DMI strength $D_{x}$ and $D_{y}$ must in opposite direction, which is $D_{x} = -D_{y}$. It is theoretically shown that, in a double layer of Fe grown in W (110) exhibits $C_{2v}$ symmetry, the antiskyrmions are stabilized with opposite sign of the DMI~\cite{gungordu2016stability}. The anisotropy interfacial DMI can be realized experimentally in ultrathin epitaxial Au/Co/W (110) system~\cite{camosi2017anisotropic}, and antiskyrmions in thin films with anisotropy DMI are realized by micromagnetic simulation~\cite{camosi2018micromagnetics}. Moreover, some theoretical investigations have shown that antiskyrmions also exist in frustrated ferromagnetic film~\cite{zhang2017skyrmion,liang2018magnetic,koshibae2016theory}.

Current induced skyrmions are investigated theoretically and experimentally~\cite{knoester2014phenomenology,jiang2017direct,iwasaki2013universal}, while the skyrmion Hall effect limits the applications of skyrmion~\cite{zhang2016magnetic,litzius2017skyrmion,kim2018asymmetric}. For antiskyrmion, the antiskyrmion Hall effect is anisotropy~\cite{huang2017stabilization}. Depending on the current direction, the Hall effect can be suppressed or enhanced, a zero antiskyrmion Hall angle can be achieved at a given current direction. However, the Joule heating induced skyrmion or antiskyrmion instability limits the application in integrated skyrmion based spintronic devices in current induced skyrmion or antiskyrmion motion. An efficient way to drive or control the skyrmion dynamics is using an electric field with lower energy consumption ~\cite{upadhyaya2015electric,zhang2015magnetic,kang2016voltage}. Skyrmions can be guided along a desired trajectory or used in skyrmion-based transistor with applying a local electric field by modifying perpendicular magnetization anisotropy~\cite{upadhyaya2015electric,song2017skyrmion}. Moreover, the skyrmion can be driven by a microwave magnetic field~\cite{moon2016skyrmion,wang2015driving} or microwave electric field~\cite{yuan2018rock,takeuchi2018selective}. However, the behaviors of an antiskyrmion under the microwave electric field are not reported.

In this work, we investigate the antiskyrmion motion under a perpendicular microwave electric field, which induces a pumping of perpendicular magnetic anisotropy (PMA), thus results in the breathing with expansion and contraction of antiskyrmion structure. By applying an in-plane magnetic field, the rotation symmetry of in-plane magnetization in antiskyrmion is broken. With the combination of these two effects, the antiskyrmion propagates with a trochoidal-like trajectory. We investigate the effect of the damping constant, the amplitude and frequency of oscillation PMA and the strength of in-plane magnetic field on the antiskyrmion motion. Moreover, using a modified Thiele equation, we analyze the skyrmion and antiskyrmion Hall angle under in-plane magnetic field with different directions, we find that the analysis is agreed with the simulation results perfectly.
% Put \label in argument of \section for cross-referencing
%\section{\label{}}
\section{Simulation model}
\begin{figure}
	\begin{center}
		\includegraphics[width=8cm]{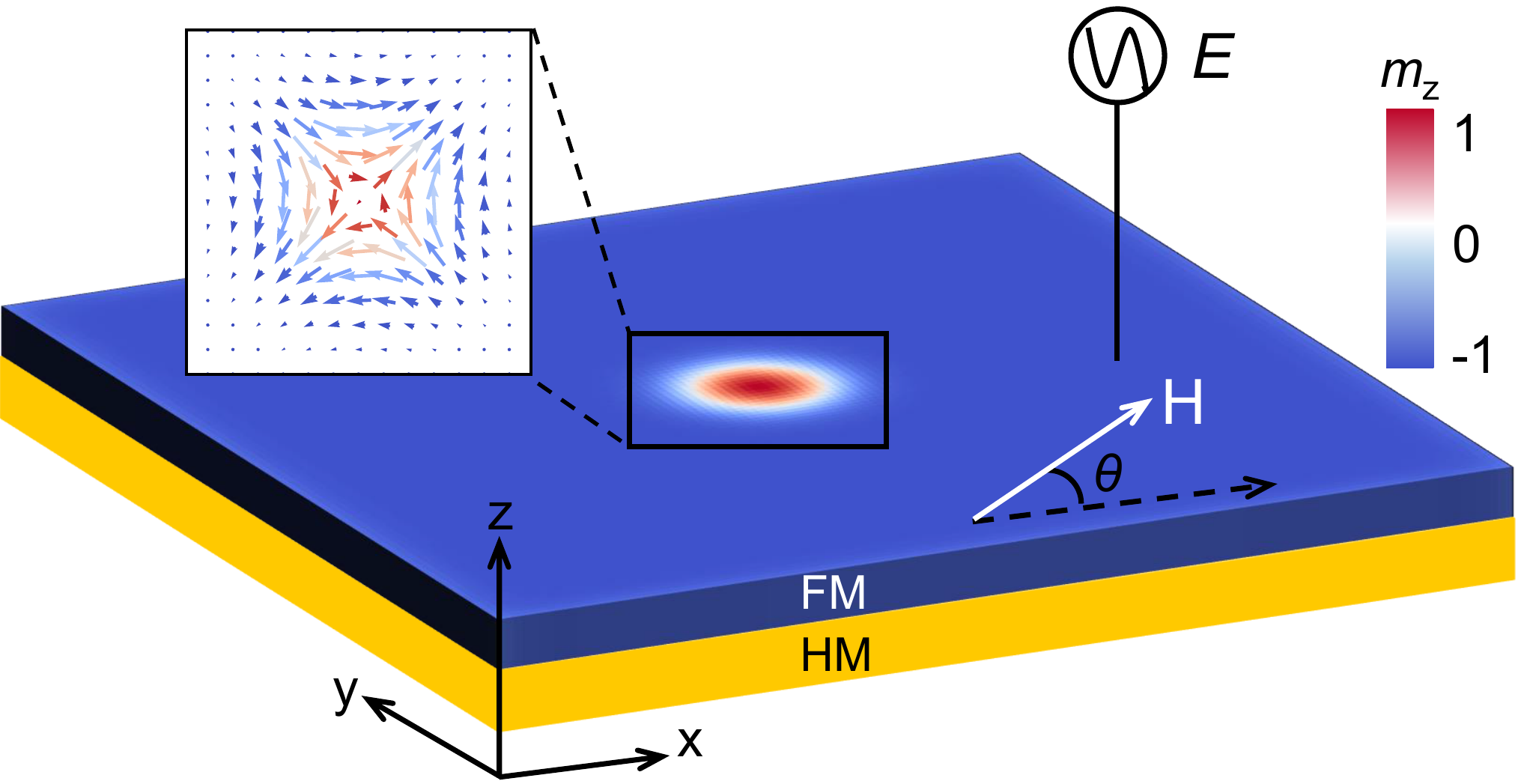} \caption{
			Schematic illustration of simulation system, which consists of a ferromagnetic (FM) layer and a heavy metal (HM) layer, which produces anisotropy DMI. An antiskyrmion is stabilized in the center of FM layer. The colors represented $m_z$ is shown in the right part. An in-plane magnetic field is applied with the angle $\theta$ with respect to $x$ axis. Microwave electric field is applied along $z$-direction which is perpendicular to the film. Inset shows the antiskyrmion structure. }
		\label{fig1}
	\end{center}
\end{figure}

In the following simulation, we consider a ferromagnetic film (FM) on a substrate, which produces anisotropy DMI, as shown in Fig.~\ref{fig1}. To investigate the dynamics of  antiskyrmion, we use the micromagnetic simulation code Mumax3~\cite{vansteenkiste2014design}, which includes a modified DMI with $D_{2d}$ crystallographic type. The antiskyrmion dynamics is govern by the Landau-Lifshitz-Gilbert (LLG) equation
\begin{equation}
\frac{\partial \mathbf{m}}{\partial t}=-\gamma \mathbf{m}\times\mathbf{H}_\mathrm{eff}+\alpha\mathbf{m}\times\frac{\partial \mathbf{m}}{\partial t},
\end{equation}
where $\mathbf{m}$ is the unit vector of magnetization, $\gamma$ is the gyromagnetic ratio, $\alpha$ is Gilbert damping constant. $\mathbf{H}_\mathrm{eff} = 2A\nabla^2\mathbf{m} + 2Km_z\mathbf{e}_z + \mathbf{H}_\mathrm{DM} + \mathbf{H}_\mathrm{in} + \mathbf{H}_d$ is the effective field of the system, which consists of exchange field, perpendicular magnetic anisotropy field, DMI field, in-plane magnetic field and dipolar field. $A$ and $K$ are exchange stiffness and magnetic anisotropy pumping, respectively. $K = K_u+K_0\sin(2\pi f t)$, where $K_u$ is the PMA of the sample, $K_0$ and $f$ represent the amplitude of oscillation magnetic anisotropy and the oscillation frequency under the microwave electric field, respectively. The magnetic film is 128 nm length, 128 nm width and 0.6 nm thick, and the unit cell size is 1 nm $\times$ 1 nm $\times$ 0.6 nm. In the competition of anisotropy interfacial DMI, exchange and perpendicular magnetic anisotropy field, an antiskyrmion is stabilized in the center of the system, as shown in the inset of Fig.~\ref{fig1}. We choose the simulation parameters of our system as: $A = 15\times10^{-12} \ \mathrm{J/m}$, $K_u = 0.8\times10^6 \ \mathrm{J/m^3}$, $M_s = 580\times10^5 \ \mathrm{A/m}$, similar to the parameters used in Ref.~\cite{sampaio2013nucleation,huang2017stabilization}. Gilbert damping varies from 0.02 to 0.2, and the DMI strength is set as $D=3.0 \ \mathrm{mJ/m^2}$. Microwave electric field induced anisotropy pumping is considered that the amplitude varies from $0.1\times10^5 \ \mathrm{J/m^3}$ and $f$ is in the range of $0$ to $40 \ \mathrm{GHz}$.

\section{Comparison of antiskyrmion and skyrmion trajectories}
\begin{figure}
	\begin{center}
		\includegraphics[width=8cm]{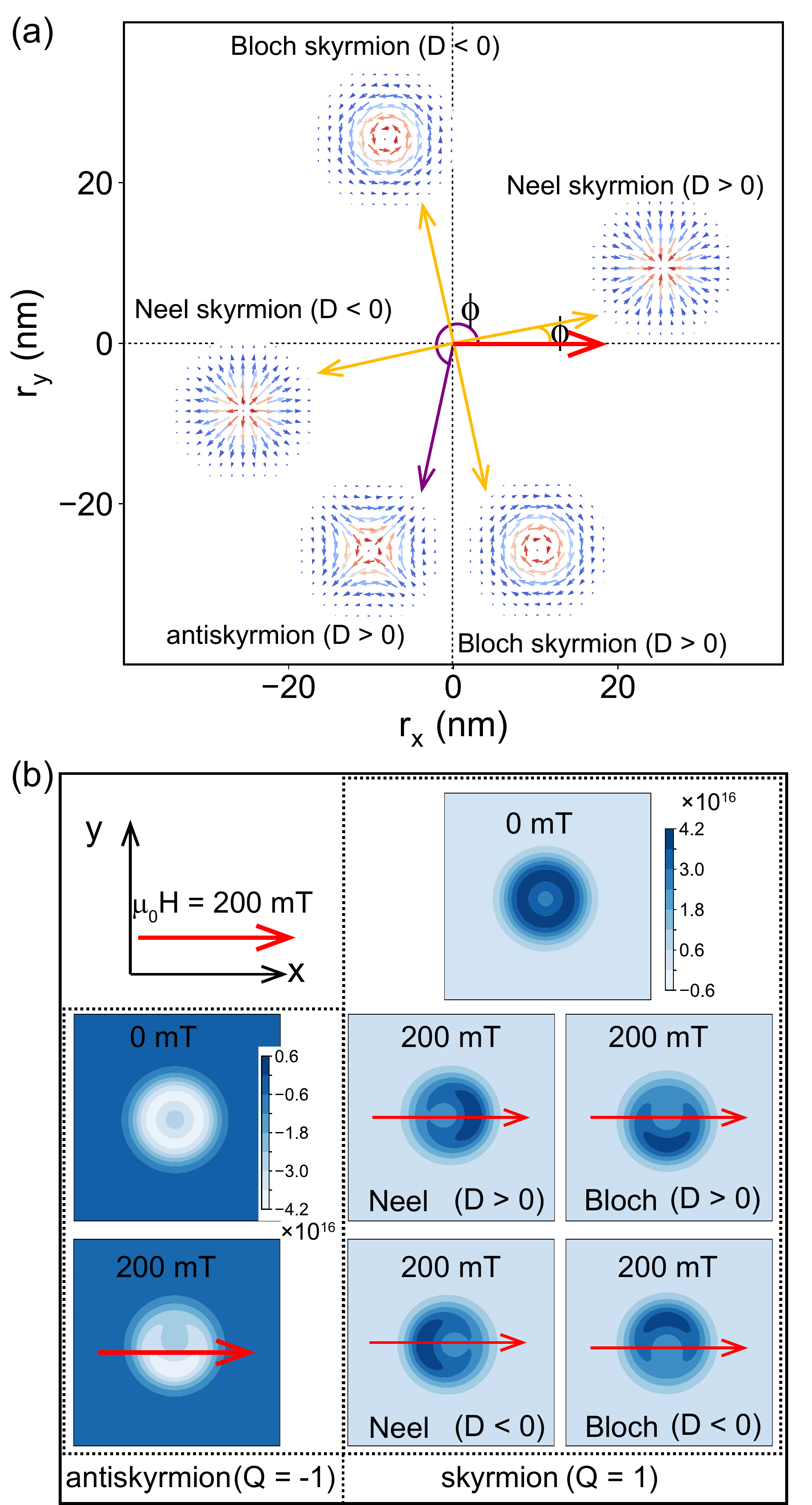} \ 
		\caption{(a) Antiskyrmion and skyrmion topological trajectories, $\phi$ is the angle between the propagation direction (purple and yellow arrows) and in-plane magnetic field (red arrow) along $x$-direction. Insets are corresponding magnetization structures. (b) Antiskyrmion and skyrmion topological density distribution $q(x, y)$ with and without magnetic field, where the magnetic field in $x$-axis is 200 mT.}
		\label{fig2}
	\end{center}
\end{figure}

First, we compare the antiskyrmion trajectory with four kinds of skyrmions propagation trajectories under an in-plane magnetic field and a perpendicular microwave electric field with the same DMI strength $|D| = 3 \ \mathrm{mJ/m^3}$, where four kinds of skyrmions are Bloch skyrmions with opposite chirality ($D>0$ and $D<0$) and N$\acute{\mathrm{e}}$el skyrmions with opposite chirality ($D>0$ and $D<0$). The strength of magnetic field applied along $x$ direction is set as 200 mT, amplitude of anisotropy pumping is $K_0 = 0.3\times10^5 \ \mathrm{J/m^3}$ and frequency $f = 14 \ \mathrm{GHz}$, the damping constant is $\alpha = 0.02$ as well. Fig.~\ref{fig2} (a) shows the trajectories of antiskyrmion and four kinds of skyrmions, the propagation angle is characterized by $\phi$, which is defined as the angle between the magnetic field and propagation direction. The results depict that the antiskyrmion propagates with $\phi = 264^\circ$. While for the N$\acute{\mathrm{e}}$el skyrmion ($D>0$), it moves with an angle $\phi = 6^\circ$. The N$\acute{\mathrm{e}}$el skyrmion with $D<0$ moves with the opposite direction of N$\acute{\mathrm{e}}$el skyrmion with $D>0$. However, the Bloch skyrmion, which induced by bulk DMI, moves with a direction perpendicular to N$\acute{\mathrm{e}}$el skyrmion. When $D>0$, the Bloch skyrmion moves with an angle $\phi = 276^\circ$, while it moves with $\phi = 96^\circ$ for $D<0$. These results reveal that, for interfacial or bulk DMI, skyrmions with opposite chiralities propagete in opposite directions. The skyrmions propagation directions are corresponding to the helicities $\gamma$, which exhibit a difference of $\pi/2$ between each other for four kinds of skyrmions~\cite{nagaosa2013topological}. Fig.~\ref{fig2} (b) shows the topological density distribution $q(x, y)$ of antiskyrmion and skyrmions with and without the in-plane magnetic field in $x$-direction. For antiskyrmion, the topological number is $Q=-1$ rather than skyrmion characterized by $Q=1$~\cite{nagaosa2013topological}, where
\begin{equation}
Q = \frac{1}{4\pi}\int \mathbf{m}\cdot (\partial_x \mathbf{m}\times\partial_y\mathbf{m})dxdy,
\end{equation}
 The topological center of antiskyrmion keeps a downward shift compared to the $q(x, y)$ distribution in 0 mT. For N$\acute{\mathrm{e}}$el skyrmion, the topological center shifted in $x$ and $-x$ directions for $D>0$ and $D<0$, respectively. While for Bloch skyrmion, the topological center displaced in $-y$ and $y$ directions for $D>0$ and $D<0$, respectively. $q(x, y)$ distribution reveals the symmetry of antiskyrmion and skyrmions directly. Applying in-plane magnetic field breaks the rotational symmetry of antiskyrmion and skyrmions. In the same microwave electric field and magnetic field, skyrmions propagation directions depend on the symmetry and chiralities. It is worth noted that for skyrmions and antiskyrmion, the propagation velocities are same under the same conditions. In the following discussion, we focus our attention on the antiskyrmion in the presence of anisotropy DMI.

\section{Antiskyrmion motion}
\begin{figure}[!htb]
	\begin{center}
		\includegraphics[width=8cm]{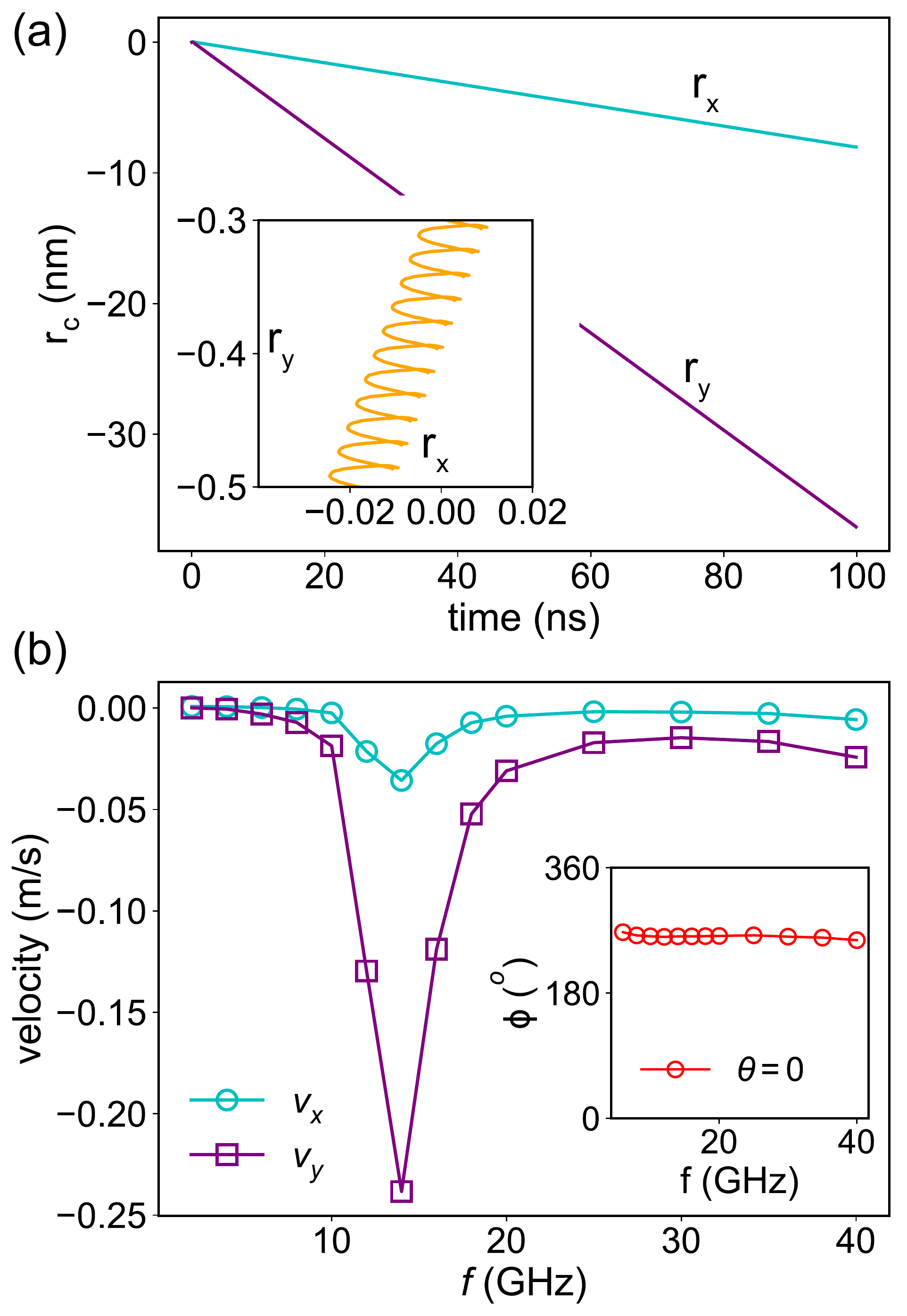} \caption{
			(a) The displacement of antiskyrmion along $x$ (cyan) and $y$ (purple) axis as a function of simulation time. Inset shows the topological trajectory of antiskyrmion. (b) Frequency dependent antiskyrmion velocities along $x$ and $y$ directions. Inset depicts the corresponding angle $\phi$.}
		\label{fig3}
	\end{center}
\end{figure}

Figure.~\ref{fig3} (a) depicts the time dependent antiskyrmion topological center $\mathbf{r}_c$ for $f = 14 \ \mathrm{GHz}$, where $\mathbf{r}_c = (r_x, \ r_y)$ is defined as~\cite{papanicolaou1991dynamics}
\begin{equation}
r_x = \frac{\int\int xq(x, y)dxdy}{\int q(x, y)dxdy}, \ r_y = \frac{\int \int yq(x, y)dxdy}{\int q(x, y)dxdy}.
\end{equation}
The result shows that antiskyrmion moves faster in $y$ direction than that in $x$ direction. Antiskyrmion motion exhibits trochoidal-like topological trajectory, where the topological center moves in a counterclockwise spiral with simulation time, as depicted in the inset of Fig.~\ref{fig3} (a). The frequency dependent antiskyrmion velocities $\mathbf{v} = (v_x, v_y)$ in $x$ and $y$ directions are shown in Fig.~\ref{fig3} (b), which reach a peak value around $f=14 \ \mathrm{GHz}$. The inset shows the related $\phi$ for different frequencies keeps a constant value about $264^\circ$, which indicates that antiskyrmion propagation direction is independent of $f$.

\begin{figure}[!htb]
	\begin{center}
		\includegraphics[width=8cm]{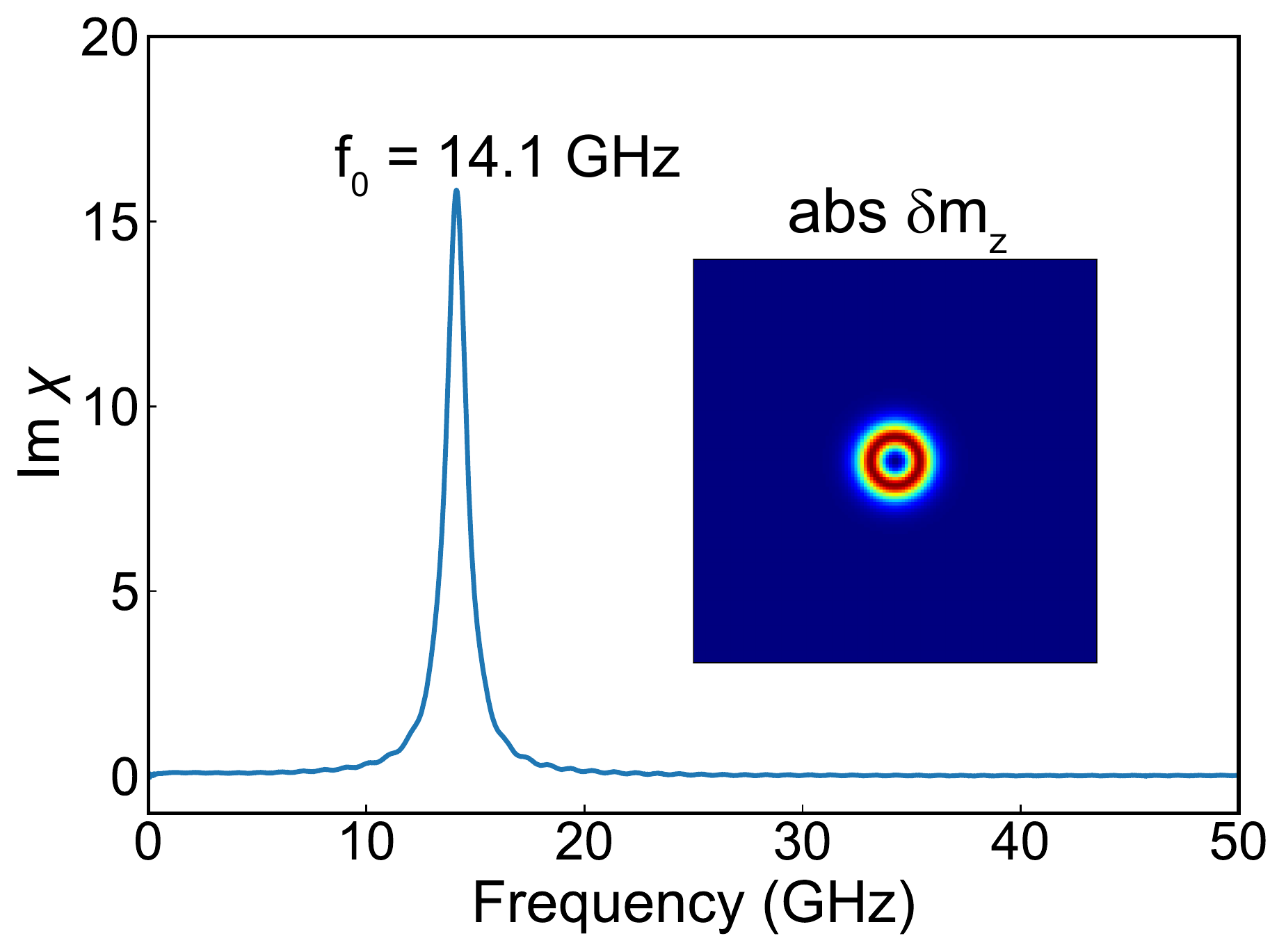} \caption{
			Imaginary part of the susceptibility spectrum of antiskyrmion in the sample. The inset shows the spatial distribution of the FFT power at eigenfrequency $f = 14.1 \ \mathrm{GHz}.$}
		\label{fig4}
	\end{center}
\end{figure}
In order to figure out why a peak antiskyrmion velocity appears in $f = 14 \ \mathrm{GHz}$, we calculate the magnetic absorption spectrum of antiskyrmion. We consider a magnetic field pulse in $z$-direction normal to the plane of system with a $sinc$ function field, $H_z(t) = H_0sinc(2\pi ft)=H_0sin(2\pi ft)/(2\pi f t)$, where $H_0 = 10 \ \mathrm{mT}$ and $f=100 \ \mathrm{GHz}$. The magnetic spectrum of simulation system is shown in Fig.~\ref{fig4} with $\alpha = 0.02$, which depicts that the resonance peak lines at $f=14.1 \ \mathrm{GHz}$. The inset is the corresponding resonance amplitude distribution, which is obtained by the fast Fourier transform (FFT) to the spatial $m_z$ oscillations of the system. The resonance peak corresponds to the breathing mode of antiskyrmion, where the antiskyrmion expands and shrinks around the antiskyrmion core with simulation time. Thus, the microwave electric field frequency corresponding to maximal antiskyrmion velocity coincides with the frequency related to eigenfrequency of the system. These results reveal that the antiskyrmion responds to the magnetic anisotropy pumping is the strongest when the frequency is in consist with eigenfrequency.
\begin{figure*}[!htb]
	\begin{center}
		\includegraphics[width=16cm]{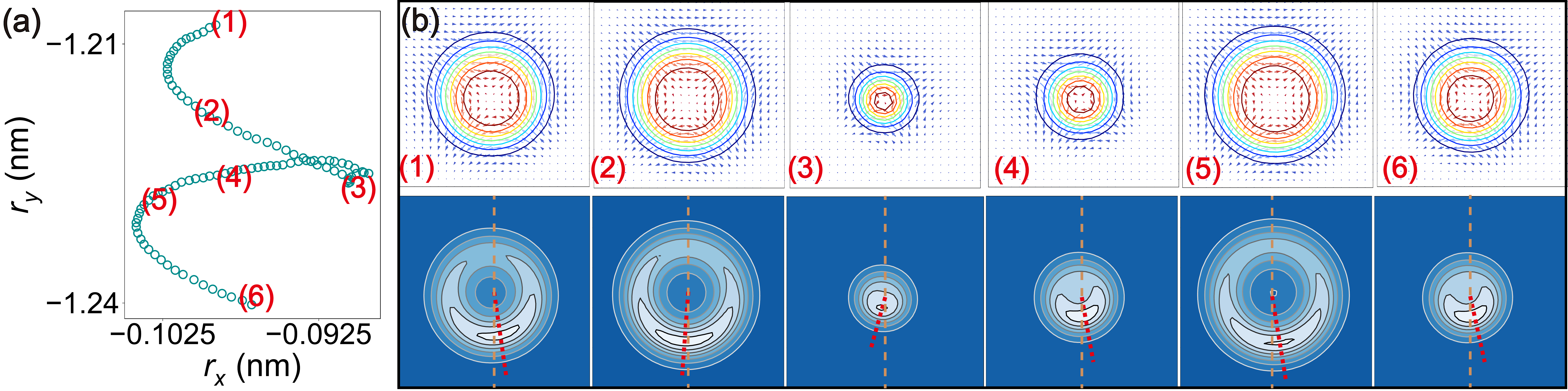} \caption{
			(a) Part of topological trajectory of antiskyrmion. (b) Corresponding antiskyrmion magnetization configurations (Upper panel) and topological density $q(x, y)$ distributions (Lower panel).}
		\label{fig5}
	\end{center}
\end{figure*}

The trochoidal motion trajectory combing breathing mode for antiskyrmion is shown in Fig.~\ref{fig5}. Fig.~\ref{fig5} (a) shows a part of the topological trajectory of antiskyrmion in Fig.~\ref{fig3} (a). In the propagation, the antiskyrmion topological center vibrates in $-x$ and $-y$ directions, the overall moving trajectory is an anticlockwise trochoidal-like motion. In the process of propagation, the antiskyrmion breathes with expansion and contraction, as shown in Fig.~\ref{fig5} (b). The order in Fig.~\ref{fig5} (b) corresponds to the numbers marked in Fig.~\ref{fig5} (a). Upper panel depicts antiskyrmion magnetization configuration, the lower panel is the corresponding $q(x, y)$ distributions. We find that the antiskyrmion size changes periodically with time. Due to the symmetry breaking caused by in-plane magnetic field, the $q(x, y)$ of antiskyrmion is different as a function of simulation time. As a result, the topological center shifted in $x$ and $y$ direction, which represented by red dashed lines in the lower panel of Fig.~\ref{fig5} (b).

\begin{figure}
	\begin{center}
		\includegraphics[width=8cm]{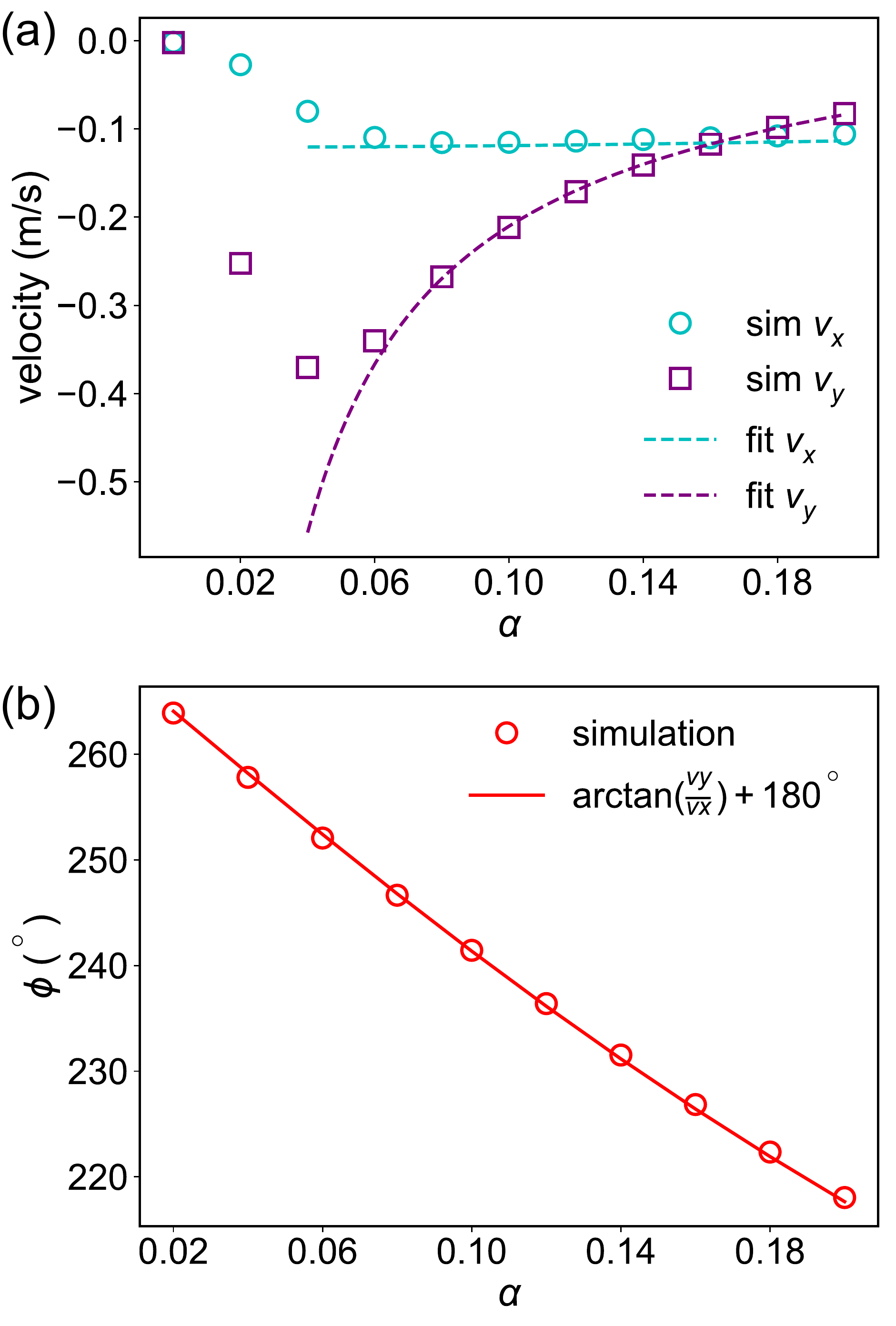} \caption{
			(a) $\alpha$ dependent antiskyrmion velocity in $x$ and $y$ directions. (b) $\phi$ as a function of $\alpha$. Points are simulation data, dashed and solid lines are fitting data.}
		\label{fig6}
	\end{center}
\end{figure}

The antiskyrmions trochoidal-like motion under microwave electric field is driven by spin waves, which are emitted by the breathing of antiskyrmion. Without applying magnetic field, the spin wave excitation is symmetric due to the rotational symmetry of antiskyrmion, thus the net driven force is zero. While applying a magnetic field along $x$-axis, the antiskyrmion symmetry is broken with the upper part of antiskyrmion wall becomes wide. At the same time, the lower part of antiskyrmion wall becomes narrow. The breaking symmetry of DW in antiskyrmion along $y$ direction induces a net driven force, which drives antiskyrmion motion with the angle $6^\circ$ respect to the $-y$ direction, as shown in Fig.~\ref{fig2} (a). The net spin wave transfer the angular momentum to antiskyrmion, which is analogy to an in-plane spin current. Here, we describe the antiskyrmion dynamics using generalized Thiele equation~\cite{thiele1973steady,sampaio2013nucleation}
\begin{equation}
\mathbf{G}\times (\mathbf{v} -\mathbf{u}^{(m)})+\mathbf{D}(\alpha \mathbf{v} - \beta \mathbf{u}^{(m)})=0,  
\end{equation}
where the boundary force is ignored. $\mathbf{G}=4\pi Q$ is the gyrovector, the sign depends on the skyrmion number $Q$. $\mathbf{D}$ is the dissipation matrix determined by the spin configurations in antiskyrmion. $\beta$ represents the misalignment of magnon polarization and local magnetization which equals to zero. $\mathbf{v}$ is the antiskyrmion propagation velocity, and $\mathbf{u}^{(m)}$ is the magnon current. The components of antiskyrmion velocity are
\begin{equation}\label{eq:34}
v_x = \frac{u_x^{(m)}-\alpha k u_y^{(m)}}{1+\alpha^2 k^2}, \  
v_y = \frac{u_y^{(m)}+\alpha k u_x^{(m)}}{1+\alpha^2 k^2}
\end{equation}
where $k = D/Q$. Fig.~\ref{fig6} (a) shows the propagation velocity of antiskyrmion as a function of Gilbert damping $\alpha$. Using the assumption in Ref.~\cite{yuan2018rock}, $v_y$ decreases with increasing $\alpha$, while $v_x$ is almost constant when $\alpha \ge 0.06$ in our system. Which suggest that $u_x^{(m)}$ is a constant, and $u_y^{(m)}$ is inversely proportional to $\alpha$.
Thus, assuming $u_x = j_x$, and $u_y = \frac{j_y}{\alpha}$, the simulation data $v_x$ and $v_y$ in Fig.~\ref{fig6} (a) can be perfectly fitted using Eq.~\ref{eq:34}, where $u_x^{(m)} = -0.0922$ and $u_y^{(m)} = -0.0235/\alpha$. The Hall angle of antiskyrmion motion is defined as $\arctan(\frac{v_y}{v_x}) = \arctan(\frac{u_y^{(m)}+\alpha k u_x^{(m)}}{u_x^{(m)}-\alpha k u_y^{(m)}})$. Fig.~\ref{fig6} (b) shows the angle between the antiskyrmion propagation and magnetic field $\phi$ as a function of $\alpha$, which equals to $\arctan{\frac{v_y}{v_x}}+180^\circ$. Using the fitting data $u_x$ and $u_y$, the simulation results can also be perfectly described, as depicted by the red line in Fig.~\ref{fig6} (b).

\begin{figure}[htb]
	\begin{center}
		\includegraphics[width=8cm]{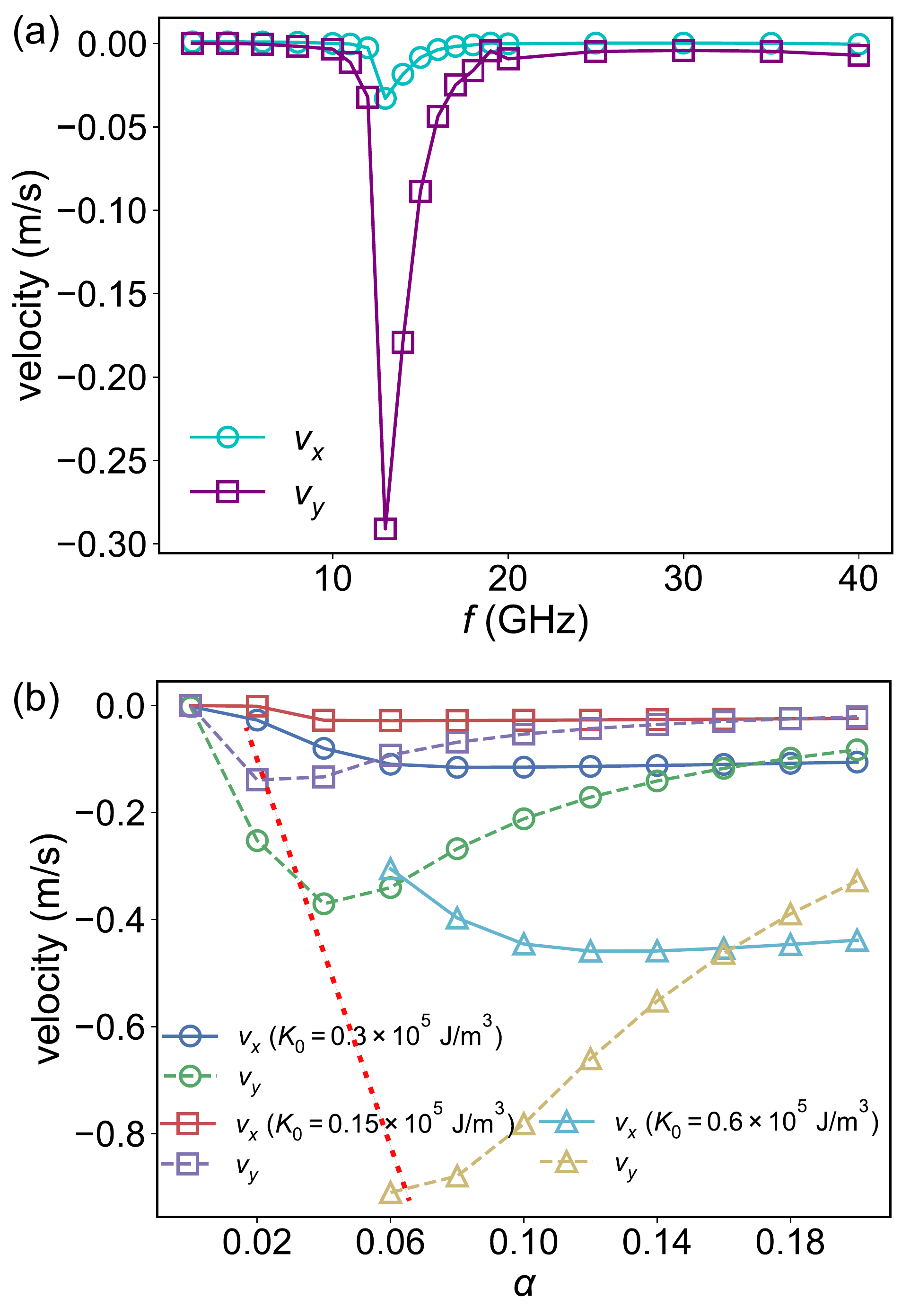} \caption{
			(a) The frequency dependent antiskyrmion velocity, where $K_0 = 0.2\times10^5 \ \mathrm{J/m^3}$, $\alpha = 0.02$. The peak frequency is 13 GHz. (b) Antiskyrmion velocity $v_x$ (solid lines) and $v_y$ (dashed lines) as a function of $\alpha$ for $K_0 = 0.15 \times10^5 \ \mathrm{J/m^3}$ (square), $\alpha$ for $K_0 = 0.3 \times10^5 \ \mathrm{J/m^3}$ (circle) and $\alpha$ for $K_0 = 0.6 \times10^5 \ \mathrm{J/m^3}$ (triangle). Dashed red line represents the shift of velocity peak.}
		\label{fig7}
	\end{center}
\end{figure}

For the purpose of indicating why the assumption in Eq.~\ref{eq:34} is not suitable for $\alpha < 0.06$, we calculate the antiskyrmion velocity for different $K_0$. Microwave electric field induced antiskyrmion breathing size is considered in the range of $R_s^{min}$ and $R_s^{max}$, with applying a positive static electric field $E_p = K_u + K_0$ or a negative static electric field $E_n = K_u - K_0$. While it is not proper with considering the antiskyrmion inertia, which will induce the oscillation amplitude of antiskyrmion out of or in the range. In Fig.~\ref{fig6} (a), the results show that the simulation data can not be fitted using Eq.~\ref{eq:34} when $\alpha < 0.06$. There exists a peak $\alpha$ which corresponds to a peak moving velocity. The antiskyrmion velocity increases with increasing $\alpha$ when $\alpha < 0.06$, which is opposite to the results where $\alpha$ is larger than 0.06. We calculate the antiskyrmion velocity as a function $f$ with $\alpha = 0.02$ and $K_0 = 0.2 \ \times 10^5 \ \mathrm{J/m^3}$, as shown in Fig.~\ref{fig7} (a). The antiskyrmion velocities reach a peak value at $f = 13 \ \mathrm{GHz}$. Compared to the results shown in Fig.~\ref{fig3} (b), the peak value exhibits a small shift about 1 GHz. The result shows that the velocity of antiskyrmion in $f = 14 \ \mathrm{GHz}$ is smaller than that in $f = 13 \ \mathrm{GHz}$ when we set $\alpha = 0.02$. Then, we change $K_0$ from $0.15\times10^5 \ \mathrm{J/m^3}$ to $0.6 \times 10^5 \ \mathrm{J/m^3}$ and the frequency is fixed to 14 GHz, the velocity as a function of $\alpha$ is shown in Fig.~\ref{fig7} (b). The results show that the peak value of antiskyrmion velocity related $\alpha$ increases with increasing $K_0$, as represented by the dashed red line in Fig.~\ref{fig7} (b). Under the same $K_0$, the peak value of antiksyrmion velocity depends on $\alpha$. While in a same $f$, the peak value related $\alpha$ exhibits a shift with increasing $K_0$. 

\begin{figure}[htb]
	\begin{center}
		\includegraphics[width=8cm]{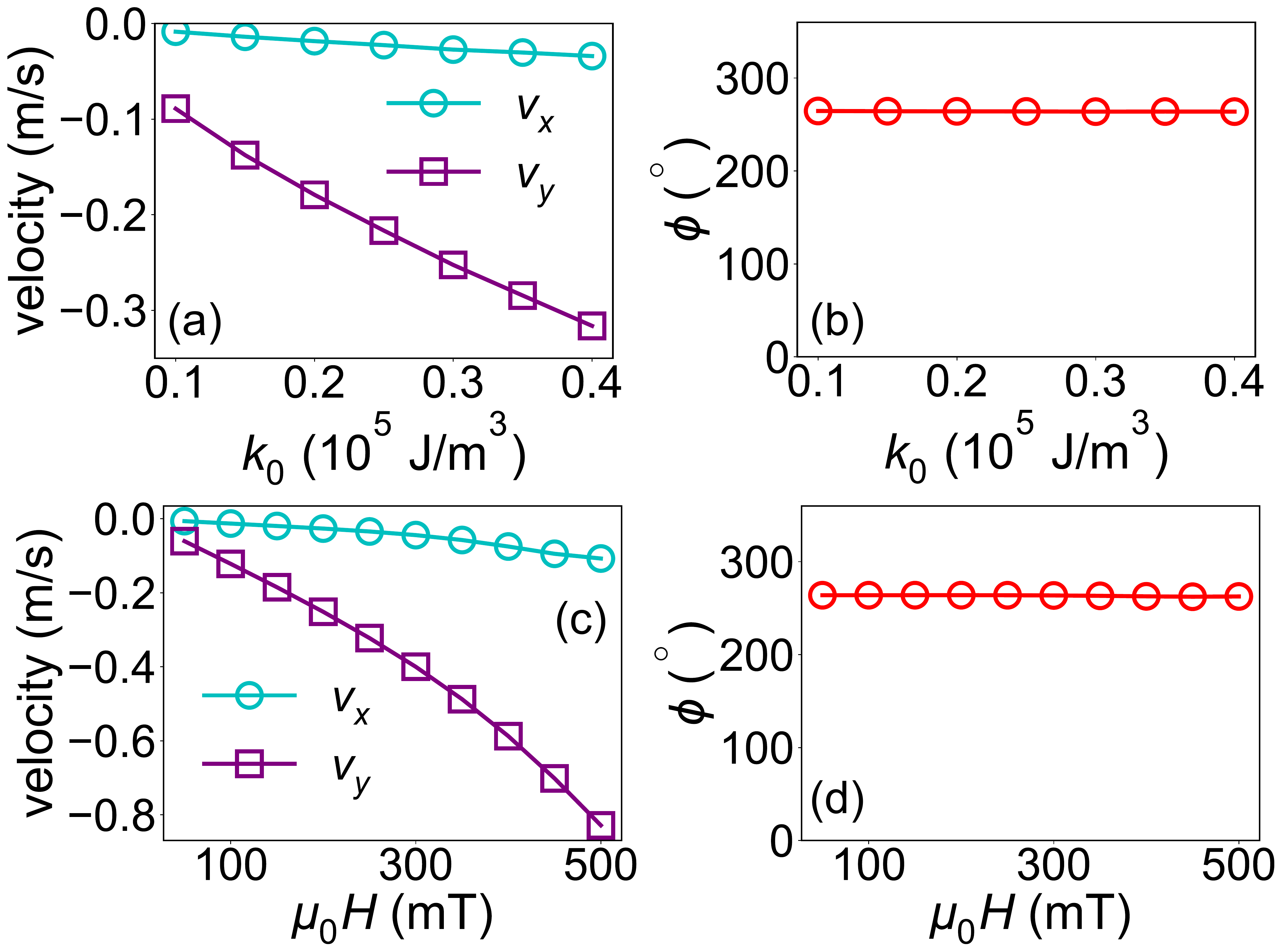} \caption{
			(a)  Antiskyrmion velocity and (b) $\phi$ as a function of $K_0$ at $\mu_0 H = 200 \ \mathrm{mT}$. (c) Antiskyrmion velocity and (b) $\phi$ as a function of $\mu_0 H$ at $K_0 = 0.3 \times 10^5 \ \mathrm{J/m^3}$.}
		\label{fig8}
	\end{center}
\end{figure}

In this part, we investigate the effect of the amplitude of anisotropy pumping $K_0$ and the strength of in-plane magnetic field on the antiskyrmion velocity and Hall angle $\phi$, where $\alpha = 0.02$ and $f = 14 \ \mathrm{GHz}$. Fig.~\ref{fig8} shows that the antiskyrmion velocity increases with increasing $K_0$ and the in-plane magnetic field. The results reveal that the spin wave emitted by the oscillation of antiskyrmion depends on $K_0$ and $\mu_0 H$. A larger pumping amplitude induces a large antiskyrmion size, thus the net magnon current emitted by antiskyrmion oscillation is larger than that in a small amplitude oscillation. Increasing magnetic field will cause a more significant rotation asymmetry of antiskyrmion, which results a larger net magnon current and the velocity increases at the same time. It is worth noted that, at a given $\alpha$, the change of amplitude and magnetic field will not influence the skyrmion hall angle with $\phi = 264^\circ$, which is the same as the effect of frequency shown in the inset of Fig.~\ref{fig3} (b). 

\section{Magnetic Field direction dependent antiskyrmion Hall effect}

\begin{figure*}[htb]
	\begin{center}
		\includegraphics[width=15cm]{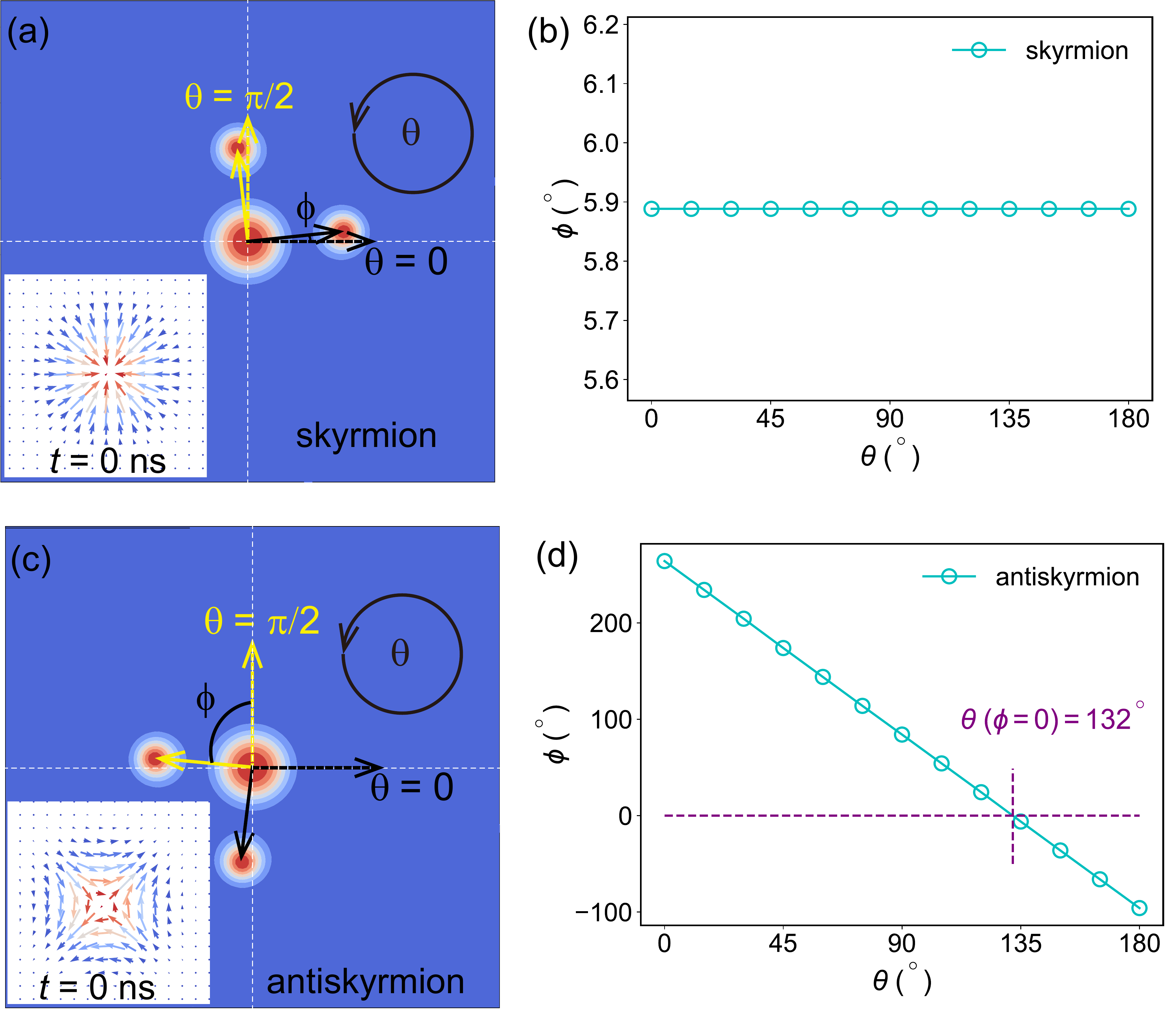} \caption{
			(a) Skyrmion motion trajectory (solid arrows) with the magnetic field direction at $\theta = 0$ (dashed black arrow) and $\theta = \frac{\pi}{2}$ (dashed yellow arrow). Inset is the skyrmion spin structure of initial state, $\phi$ is an angle denotes the skyrmion motion trajectory with respect to the direction of magnetic field. (b) Skyrmion Hall angle $\phi$ as a function of $\theta$. (c) Antiskyrmion propagation trajectory (solid arrows) with the magnetic field direction at $\theta = 0$ (dashed black arrow) and $\theta = \frac{\pi}{2}$ (dashed yellow arrow). Inset is the antiskyrmion spin structure of initial state. (d) $\phi$ for antiskyrmion as a function of $\theta$. The antiskyrmion motion trajectory and the direction of magnetic field are along the same direction when $\theta = 132^\circ$. }
		\label{fig9}
	\end{center}
\end{figure*}

The results above show that the antiskyrmion motion exhibits a Hall effect driven by magnon current under the combined action of magnetic field along $x$-axis and microwave electric field, which is the same as skyrmion propagation driven by in-plane current or spin Hall effect. In this section, we investigate the skyrmion (N$\acute{\mathrm{e}}$el with $Q=1$ and $D>0$) and antiskyrmion ($Q=-1$ and $D>0$) propagation directions with varying the direction of in-plane magnetic field. Moreover, whether there is a difference for the isotropy and anisotropy spin configurations of skyrmion and antiskyrmion as a function of magnetic field direction. Fig.~\ref{fig9} (a) and (c) show the typical propagation trajectories of skyrmion and antiskyrmion in the presence of frequency $f = 14 \ \mathrm{GHz}$, Gilbert damping $\alpha = 0.02$ and amplitude $K_0 = 0.3 \times 10^5 \ \mathrm{J/m^3}$. The direction of magnetic field is characterized by $\theta$, and the magnitude is given as 200 mT. We set $\theta = 0$, which means the magnetic field is applied along $x$-direction, the skyrmion exhibits a longitudinal motion along $x$-axis and a small transverse motion along $y$-axis. The angle between the propagation direction and the direction of magnetic field is $\phi = 5.9^\circ$, which is calculated as $\arctan(v_y/v_x)$. For antiskyrmion, it moves with a large longitudinal motion along -$y$-axis and a small transverse motion along -$x$-aixs, the transverse motion is quantified as $\phi = \arctan(v_y/v_x) + \pi$ which is $264^\circ$. When a magnetic field along $y$-axis is applied ($\theta = \pi/2$), the angle $\phi$ for skyrmion is still $5.9^\circ$, while $\phi = 84.14^\circ$ for antiskyrmion. The magnetic field direction $\theta$-dependent $\phi$ for skyrmion and antiskyrmion are depicted in Fig.~\ref{fig9} (b) and (d), respectively. $\phi$ for skyrmion is insensitive to field direction, which is in consist with the skyrmion Hall effect driven by current or spin Hall effect~\cite{jiang2017direct,woo2016observation}. However, $\phi$ depends on the direction of magnetic field and decreases with increasing $\theta$ for antiskyrmion. The field direction dependent directional motion phenomenon reveals that the skyrmion motion described by skyrmion Hall angle is not applicable for antiskyrmion motion under microwave electric field and magnetic field. 

In Fig.~\ref{fig6}, we have obtained that the magnon current in $x$-direction is $u^{as}_x = -0.0922$, which $u^{as}_y = -0.0235 / 0.02$ in $y$-direction in the same microwave electric field and the magnetic field 200 mT along $x$-axis with $\alpha  = 0.02$. Thus, the net magnon current generated by the breathing of antiskyrmion is 
\begin{equation}
u = \sqrt{u_x^2 + u_y^2}.
\end{equation}
Under the same magnetic field and microwave electric field, the net magnon current $u$ generated by microwave and in-plane magnetic field is insensitive to the differences of the isotropy or anisotropy spin configurations for skyrmion and antiskyrmion. After comparing with the propagation direction of antiskyrmion, the magnon current components along $x$ and $y$ directions for skyrmion are $u^s_x = -u^{as}_y$ and $u^s_y = -u^{as}_x$. The skyrmion Hall angle can be calculated as $\phi_0 = \arctan(v_y/v_x) = 5.935^\circ$, which is very close to the simulation results $5.9^\circ$. Due to the isotropy spin configuration of skyrmion, the generated magnon current always exhibits an angle $\eta_s = \arctan{(u^s_y/u^s_x)} = 4.478^\circ$ with the direction of magnetic field. Thus the skyrmion Hall angle $\phi$ keeps a fixed value with the direction of magnetic field. In contrast to skyrmion with isotropy spin texture, the antiskyrmion have anisotropy in-plane magnetizations, thus the field-dependent symmetry breaking depends on the anisotropy in-plane magnetizations. In the following, we explore the anisotropy response of antiskyrmion motion for different directions of in-plane magnetic field. The field direction dependent antiskyrmion motion can be understood by a modified Thiele equation with introducing field direction dependent magnon current
\begin{equation}
 \mathbf{G}\times \mathbf{v} + \mathbf{D} \alpha \mathbf{v} = \mathbf{C} \mathbf{F},
\end{equation}
where $\mathbf{F} = \mathbf{G}\times \mathbf{u}^{(m)}$ and $\mathbf{C}$ is a matrix that describes the spin configurations according to the symmetry of DMI, due to the reason that the generated magnon current direction depends on the direction of magnetic field. Thus we get that
\begin{equation}
u_x = u\cos(\eta + \theta), \ u_y = u\sin(\eta + \theta)
\end{equation}
where $\eta$ characterizes the direction of magnon current respect to $x$-axis with $\theta = 0$, which is $\eta_s = 4.478^\circ$ for skyrmion and $\eta_{as} = 85.522^\circ$ for antiskyrmion. For skyrmion isotropy DMI, $D_x = D_y$, we get 
\begin{equation}
C_{ij} = \left[\begin{matrix}
C_{xx} & 0 \\
0 & C_{yy}
\end{matrix}
\right] = C\left[\begin{matrix}
1 & 0 \\
0 & 1
\end{matrix}
\right],
\end{equation}
The ratio between the velocity along $y$-axis and $x$-axis is 
\begin{equation}
\begin{aligned}
\frac{v_y}{v_x} = & \frac{C_{yy} u\sin(\eta_s + \theta)+\alpha k C_{xx}u\cos(\eta_s + \theta)}{C_{xx}u\cos(\eta_s+ \theta)-\alpha kC_{yy}u\sin(\eta_s + \theta)} \\
= & \tan(\phi_0 + \theta),
\end{aligned}
\end{equation}
where $\phi_0$ is the angle of skyrmion motion respect to field direction with $\theta = 0$. Thus, $\phi = \arctan(\frac{v_y}{v_x}) - \theta=\phi_0$, which depicts that the skyrmion moving direction is independent of the field direction. This is in consist with our simulation.

While for antiskyrmion with anisotropy DMI, $D_x = -D_y$, we get 
\begin{equation}
C_{ij} = \left[\begin{matrix}
C_{xx} & 0 \\
0 & -C_{yy}
\end{matrix}
\right] =C\left[\begin{matrix}
1 & 0 \\
0 & -1
\end{matrix}
\right],
\end{equation} 
The corresponding ratio of $v_y$ and $v_x$ is
\begin{equation}
\begin{aligned}
\frac{v_y}{v_x} = & -\frac{C_{yy} u\sin(\eta_{as}+ \theta)+\alpha k C_{xx}u\cos(\eta_{as} + \theta)}{C_{xx}u\cos(\eta_{as} + \theta)+\alpha kC_{yy}u\sin(\eta_{as} + \theta)} \\
= & \tan(\phi_0 - \theta),
\end{aligned}
\end{equation}
where $\phi_0 = 265.5^\circ$ is the angle of antiskyrmion motion with respect to $x$-axis when $\theta = 0$. Thus, $\phi = \arctan(\frac{v_y}{v_x}) - \theta=\phi_0 - 2\theta$. Which depicts that the antiskyrmion motion trajectory depends on the field direction $\theta$ with a slpoe of -2, which is in consistent with our simulation results that $\phi$ decreases with increasing $\theta$ (Fig.~\ref{fig9} (d)). The combined action of magnetic field and microwave electric field introduces a method to drive antiskyrmion motion, while magnetic fields direction dependent antiskyrmion motion give rise to an opportunity to control the trajectory antiskyrmion. The angle $\phi$ vanishes when the magnetic field is applied along $\theta = \phi_0/2=132.7^\circ$, which corresponds to the simulation result shown in Fig.~\ref{fig9} (d) with $\theta (\phi = 0) = 132^\circ$. This allow for an antiskyrmion motion along the direction of magnetic field, which provides a unique method in the application of antiskyrmion-based spintronic devices.

\section{Conclusion}
In summary, we have shown that the antiskyrmion exhibits a trochoidal-like motion under the microwave electric field in the presence of in-plane magnetic field. The anisotropy pumping induces antiskyrmion breathing with emitting spin waves, while the net magnon current is zero due to the rotation symmetry. Applying a in-plane magneitc field breaks the rotation symmetry, and a net magnon current is generated which drives antiskyrmion motion. We find that the antiskyrmion velocity reaches a maximum value with the microwave electric field frequency corresponding to the resonance frequency of the sample. In a small $\alpha$, the peak value for maximum velocity exhibit a small shift. Moreover, increasing the amplitude of the anisotropy pumping and magnetic field will increase the antiskyrmion moving velocity. Interestingly, the angle between the propagation direction and the direction of magnetic field $\phi$ is independent of magnitudes of frequency, amplitude and the strength of in-plane magnetic field. We show that the antiskyrmion moving direction depends on the direction of in-plane magnetic field, the angle $\phi$ exhibits an anisotropy response. Using the modified Thiele equation with introducing the symmetry of DMI, we analysis this phenomenon perfectly. Our results depict that the antiskyrmion motion with lower consumption driving method may be potential in the future spintronic devices. 
\section*{Acknodledgement}
This work is supported by National Science Fund of China (Grants No. 11574121 and No. 51771086)

\bibliography{references}

\end{document}